\documentclass[journal,10pt]{IEEEtran}

\usepackage{acro}
\acsetup{first-style=short}
\usepackage{ifthen}

\newboolean{printTODO}
\newboolean{printAbstract}

\setboolean{printTODO}{false}
\setboolean{printAbstract}{true}

\makeatletter
\def\markboth#1#2{\def\leftmark{\@IEEEcompsoconly{\sffamily}\MakeUppercase{\protect#1}}%
\def\rightmark{\@IEEEcompsoconly{\sffamily}\MakeUppercase{\protect#2}}}
\makeatother

\usepackage[american]{babel}
\usepackage[utf8]{inputenc} 

\usepackage{grffile}

\usepackage{etex}

\usepackage[hidelinks]{hyperref}

\ifCLASSINFOpdf
  \usepackage[pdftex]{graphicx}
	\usepackage{epstopdf}
  \graphicspath{{graphics/}}
  \DeclareGraphicsExtensions{.eps,.pdf,.jpeg,.jpg,.png}
\else
  \usepackage[dvips]{graphicx}
  \graphicspath{{graphics/}}
  \DeclareGraphicsExtensions{.eps}
\fi

\usepackage[cmex10]{amsmath}
\usepackage{array}

\usepackage{eqparbox}

\usepackage[caption=false,font=footnotesize]{subfig}
\usepackage{subfig}

\usepackage{stfloats}

\usepackage{url}

\ifthenelse{\boolean{printTODO}}
{
	\usepackage{tikz}
	
	\usetikzlibrary{circuits}
	\usetikzlibrary{circuits.ee.IEC}
	\usepackage[colorinlistoftodos,textsize=scriptsize]{todonotes}
	\setlength{\marginparwidth}{1.3cm}
}
{
	\usepackage[disable,colorinlistoftodos]{todonotes}
}

\usepackage[backend=biber,style=ieee,sorting=none,doi=true,isbn=false,url=false,minbibnames=1,maxbibnames=1,mincitenames=1,maxcitenames=1]{biblatex}
\uspunctuation

\usepackage[alsoload=astro, alsoload=hep]{siunitx}
\usepackage{placeins} %

\usepackage{isotope}

\usepackage{xspace}
\makeatletter
\let\@autoref=\autoref
\renewcommand*{\autoref}[2][]{\ifthenelse{\equal{#1}{}}{\@autoref{#2}}{\hyperref[#1]{\begin{NoHyper}\@autoref{#2}~\subref{#1}\end{NoHyper}}}\xspace}
\makeatother

\usepackage{epstopdf}

\usepackage{verbatim}

\usepackage{booktabs}
\usepackage{xcolor,colortbl}
\usepackage{tabularx}
\newcolumntype{C}[1]{>{\centering\arraybackslash}m{#1}}
\newcolumntype{R}[1]{>{\raggedleft\arraybackslash}m{#1}}

\usepackage{multirow}

\DeclareFieldFormat{sentencecase}{\MakeSentenceCase{#1}}
\renewbibmacro*{title}{%
  \ifthenelse{\iffieldundef{title}\AND\iffieldundef{subtitle}}
  {}
  {\ifthenelse{\ifentrytype{article}\OR\ifentrytype{inbook}%
      \OR\ifentrytype{incollection}\OR\ifentrytype{inproceedings}%
      \OR\ifentrytype{inreference}}
    {\printtext[title]{%
        \printfield[titlecase]{title}%
        \setunit{\subtitlepunct}%
        \printfield[titlecase]{subtitle}}}%
    {\printtext[title]{%
        \printfield[titlecase]{title}%
        \setunit{\subtitlepunct}%
        \printfield[titlecase]{subtitle}}}%
    \newunit}%
  \printfield{titleaddon}
}

\usepackage{tikz}

\usepackage{adjustbox}

\newcommand{\tikzsubfig}[3]
{
	\subfloat{
		\label{#3}
		\begin{tikzpicture}
		\node[anchor=south west,inner sep=0] (image) at (0,0) {\includegraphics[#2]{#1}};
		\begin{scope}[x={(image.south east)},y={(image.north west)}]
		\node[font=\small] at (0.06, 0.06) {\subref{#3}};
		\end{scope}
		\end{tikzpicture}
	}
}

\bibliography{bibliography.bib}

\DeclareSIUnit\percentpoint{pp}
\DeclareSIUnit\photoelectron{p.e.}
\DeclareSIUnit\gigabit{Gbit}
\DeclareSIUnit\kilocountspersecond{kcps}
\DeclareSIUnit\mWperchannel{mW/channel}

\DeclareAcronym{P0}{
	short = $P_{0}$ ,
	long  = power consumption constant,
	sort  = P0,
	class = nomencl
}

\DeclareAcronym{dPdiscsfbias}{
	short = $dP_{\mathsf{disc\_sf\_bias}}$ ,
	long  = power consumption change due to $\mathsf{disc\_sf\_bias}$,
	sort  = dPdiscsfbias,
	class = nomencl
}

\DeclareAcronym{dPfeib2}{
	short = $dP_{\mathsf{fe\_ib2}}$ ,
	long  = power consumption change due to $\mathsf{fe\_ib2}$, 
	sort  = dPfeib2,
	class = nomencl
}

\DeclareAcronym{dPfeib1}{
	short = $dP_{\mathsf{fe\_ib1}}$ ,
	long  = power consumption change due to $\mathsf{fe\_ib1}$,
	sort  = dPfeib1,
	class = nomencl
}

\DeclareAcronym{Uoff}{
	short = $U_{\mathsf{off}}$ ,
	long  = mean offset of applied bias voltage,
	sort  = Uoff,
	class = nomencl
}

\DeclareAcronym{Ubias}{
	short = $U_{\mathsf{bias,set}}$ ,
	long  = bias voltage,
	sort  = Ubias,
	class = nomencl
}

\DeclareAcronym{Ucor}{
	short = $U_{\mathsf{cor}}$ ,
	long  = bias voltage corrected for voltage offset,
	sort  = Ucor,
	class = nomencl
}

\DeclareAcronym{Ubd}{
	short = $\bar{U}_{\mathsf{BD}}$ ,
	long  = mean breakdown voltage,
	sort  = Ubd,
	class = nomencl
}

\DeclareAcronym{Ucorrel}{
	short = $U_{\mathsf{cor,rel}}$ ,
	long  = relative offset-corrected overvoltage,
	sort  = Ucorrel,
	class = nomencl
}

\DeclareAcronym{f0}{
	short = $f_{0}$ ,
	long  = power consumption parameter,
	sort  = f0,
	class = nomencl
}

\DeclareAcronym{a0}{
	short = $a_{0}$ ,
	long  = power consumption parameter,
	sort  = a0,
	class = nomencl
}

\DeclareAcronym{a1}{
	short = $a_{1}$ ,
	long  = power consumption parameter,
	sort  = a1,
	class = nomencl
}

\DeclareAcronym{a2}{
	short = $a_{2}$ ,
	long  = power consumption parameter,
	sort  = a2,
	class = nomencl
}

\DeclareAcronym{b0}{
	short = $b_{0}$ ,
	long  = power consumption parameter,
	sort  = b0,
	class = nomencl
}

\DeclareAcronym{b1}{
	short = $b_{1}$ ,
	long  = power consumption parameter,
	sort  = b1,
	class = nomencl
}

\DeclareAcronym{b2}{
	short = $b_{2}$ ,
	long  = power consumption parameter,
	sort  = b2,
	class = nomencl
}

\usepackage{multirow}

\begin{document}

\newcommand{\myTitle}{Investigation of the Power Consumption of the PETsys TOFPET2 ASIC}

\title{{\LARGE \myTitle}}

\author{Vanessa~Nadig$^{1}$\href{https://orcid.org/0000-0002-1566-0568}{\includegraphics[height=0.75em]{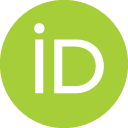}},
        Bjoern~Weissler$^{1,2}$\href{https://orcid.org/0000-0003-1119-785X}{\includegraphics[height=0.75em]{orcid_icon}},
        Harald~Radermacher$^{1}$\href{https://orcid.org/0000-0002-8603-1750}{\includegraphics[height=0.75em]{orcid_icon}},
        Volkmar~Schulz$^{1,2,3,4}$\href{https://orcid.org/0000-0003-1341-9356}{\includegraphics[height=0.75em]{orcid_icon}},
		and~David~Schug$^{1,2}$\href{https://orcid.org/0000-0002-5154-8303}{\includegraphics[height=0.75em]{orcid_icon}}%
\thanks{$^1$Department of Physics of Molecular Imaging Systems, Institute for Experimental Molecular Imaging, RWTH Aachen University, Aachen, Germany;
$^2$Hyperion Hybrid Imaging Systems GmbH, Pauwelsstrasse 19, 52074 Aachen, Germany;
$^3$III. Physikalisches Institut B, Otto-Blumenthal-Straße, 52074 Aachen, Germany;
$^4$Fraunhofer Institute for Digital Medicine MEVIS, Forckenbeckstrasse 55, Aachen Germany}%
\thanks{Corresponding author: \textcolor{black}{\url{vanessa.nadig@pmi.rwth-aachen.de}}}%
\thanks{This project has received funding from the European Union’s Horizon 2020 research and innovation programme under grant agreement No 667211 \protect\includegraphics[height=0.75em]{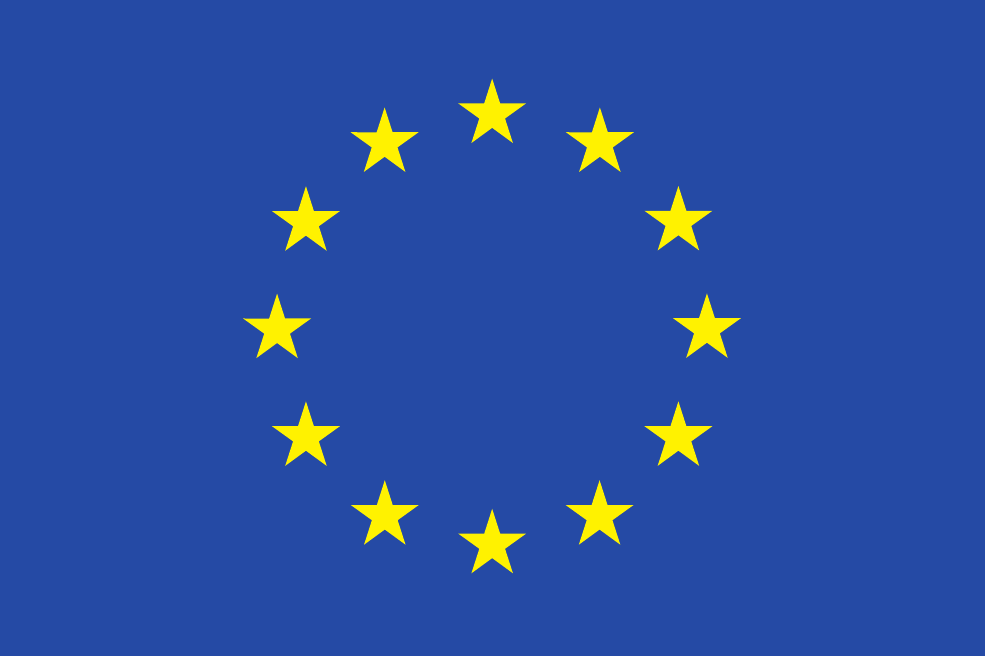}.}
}

\markboth{Accepted for publication in: IEEE TRPMS - Special Issue PSMR 2019 - 8th conference on PET/MR and SPECT/MR}%
{Nadig \MakeLowercase{\textit{et al.}}: \myTitle}

\maketitle

\ifthenelse{\boolean{printAbstract}}
{
\begin{abstract}

In state-of-the-art positron emission (PET) tomography systems, application-specific integrated circuits (ASICs) are commonly used to precisely digitize the signals of analog silicon photo-multipliers (SiPMs).
However, when operating PET electronics in a magnetic resonance (MR) system, one faces the challenge of mutual interference of these imaging techniques.
To prevent signal deterioration along long analog signal lines, PET electronics with a low power consumption digitizing the signals close to the SiPMs are preferred.
In this study, we evaluate the power consumption of the TOFPET2 ASIC.
Its power consumption ranges from \SIrange{3.6}{7.2}{\mWperchannel} as a function of the input stage impedance and discriminator noise settings.
We present an analytical model allowing to compute the power consumption of a given ASIC configuration.
The input stage impedance and discriminator noise have an impact on the coincidence resolution time, energy resolution, and photon trigger level.
Since the TOFPET2 ASIC delivers state-of-the-art performance with a power consumption similar or even lower than other ASICs typically used for PET applications, it is a favorable candidate to digitize the signals of SiPMs in future simultaneous PET/MR systems.
\end{abstract}
}{}

\section{Introduction}

\IEEEPARstart{I}{n} positron emission tomography (PET), radioactive tracer molecules are injected into the patient's body.
The tracer molecules undergo a $\beta^{+}$-decay resulting in the emission of a positron which annihilates with an electron in the surrounding tissue.
Two $\gamma$-photons released back-to-back by the electron-positron annihilation with an energy of \SI{511}{\keV} each are detected by a ring-shaped array of $\gamma$-detectors surrounding the patient \cite{surti2016advances,vandenberghe2016RecentDevelopments,phelps2006pet}.
Based on these so-called coincidence events, PET is used as a functional imaging technique in oncology, neurology and cardiology \cite{weber2003PETLungCancer,marcus2014BrainPET,nakazato2013MyocardialPerfusion}.\\
Common state-of-the-art PET systems employ scintillators fabricated of lutetium-(yttrium-)oxyorthosilicate doped with cerium (L(Y)SO) converting incident $\gamma$-photons into optical photons and analog silicon-photomultipliers (SiPMs) as photo-detectors, which have come to replace previously used avalanche photo-diodes (APDs) during the past years \cite{bisogni2016SiPMs}.
An SiPM consists of several thousand single-photon avalanche diodes (SPADs) which are connected in parallel. 
The term SPAD refers to avalanche photo-diodes (APDs), which are operated in Geiger mode. 
The SPADs break down and generate an analog pulse when hit by an optical photon. 
The signal sum is a measure for the number of detected photons. 
The timestamp of the detected $\gamma$-interaction can be determined from the first optical photons detected by the SiPM.
Passive quenching of the self-sustaining avalanche resulting from a diode breakdown and thus, resetting the diode is achieved by a serial high-ohmic resistance per individual SPAD.
Using time-of-flight (TOF) information, PET systems are capable of resolving the difference in arrival times of the two $\gamma$-photons of a coincidence event down to the order of few hundred picoseconds.
This allows to localize the annihilation event more precisely, which can be exploited during image reconstruction leading to a better signal-to-noise ratio (SNR) of the image of the activity distribution \cite{gundacker2014ToFGain,surti2015ToFGain,vandenberghe2016RecentDevelopments,surti2016advances}.
State-of-the-art clinical systems reach coincidence resolution times (CRTs) down to \SI{214}{\pico\second} -- a benchmark set by the Siemens Biograph Vision PET/CT system \cite{BiographVisionDataSheet}.
In benchtop experiments, much lower CRTs down to \SI{58}{\pico\second} are possible \cite{gundacker2019CTR58ps,lecoq2017PushingToFLimits}.\\
For SiPM readout in PET applications, the precise digitization of event timestamps and energies is typically achieved by employing application-specific integrated circuits (ASICs).
These ASICs typically support 8 to 64 data channels \cite{rolo2012TOFPETASIC,chen2014DedicatedReadoutASIC,corsi2009ASICdevelopment,fischer2009PETA,shen2012STiC,ahnen2018STiC,fleury2013Petiroc,sacco2013PETA4,ahmad2015Triroc,ahmad2016Triroc}. 
The time binning of the employed time-to-digital converters (TDC) can range from \SIrange{20}{50}{\pico\second} \cite{ahmad2015Triroc,stankova2015STIC3,sacco2013PETA4,rolo2013TOFPETASIC}.
The energy of the signal can either be measured by a time-over-threshold method (tot-mode) \cite{chen2014DedicatedReadoutASIC,orita2017CurrentToTASIC,cela2018FlexToT,gomez2019HRFlexToT} or via signal integration (qdc-mode), e.g., as applied for the Weeroc, PETA and TOFPET ASIC series \cite{ahmad2015Triroc,ahmad2016Triroc,fleury2017PETIROC2A,fischer2006MultiChannelReadout,fischer2009PETA,piemonte2012PerformancePETA3ASIC,sacco2013PETA4,sacco2015PETA5,schug2017PETA5,bugalho2017ExperimentalResultsTOFPET2,rolo2012TOFPETASIC,PETsysTOFPET1DataSheet,PETsysTOFPET2DataSheet,callier2012Easiroc}.
The measurement can be linear for integrating charges up to \SIrange{2000}{3000}{\photoelectron} (photo-electrons) \cite{ahmad2016Triroc,callier2012Easiroc,fleury2013Petiroc}.\\
When integrating TOF-PET and magnetic resonance imaging (MRI) in one hybrid system, one faces the need of a compact infrastructure designed for the only restricted space inside the MR bore as well as problems of dissipation and mutual interference.
These can result in performance degradation for both imaging modalities and, thus, need to be evaluated \cite{wehner2015MRcompatibilityassessment,vandenberghe2015PETMRIchallenges,schug2015tofring}.
In addition, one has to take into account the power supply that is required by the high-performance PET electronics.
Connecting the PET electronics to the SiPMs via long cables from outside to prevent space problems inside the MR system potentially leads to a performance degradation due to a loss of the SiPM signal quality, e.g., as specified for the use of the TOFPET2 ASIC \cite{TOFPET2ReadoutSystemHardwarUserGuide}.
The signal quality is deterioated by the increased inductance and impedance on the signal line.
The long analog signal lines would additionally call for sophisticated shielding to avoid a distortion of the transmitted signals by the dynamic magnetic fields of the MR system.
An early digitization close to the SiPM should therefore be considered.\\ %
To this end, the power for the PET electronics has to be provided via circuitry inside the MR bore. 
The MR-environment puts a lot of constraints on the selection of the power supply electronics, complicating the use of switched mode power supplies (SMPS).
Hence, linear voltage regulators are often used to provide the final supply voltage for the PET electronics, whose design as well as the required infrastructure benefits from low-power PET electronics \cite{dueppenbecker2016PETMRIdSiPM,weissler2014analogPETMR}.
For those reasons, PET electronics with a low power consumption are favored for system integration to overcome the aforementioned effects.\\
This study aims to characterize the power consumption of the TOFPET2 ASIC, which was released by PETsys Electronics S.A. in 2017 \cite{PETsysWebPage}.
In addition, the possible impact of the power consumption configuration on the ASIC performance is evaluated to further assess the system applicability of the TOFPET2 ASIC.

\section{Materials}

\begin{figure}[t]
	\centering
	\includegraphics[height=0.9\columnwidth]{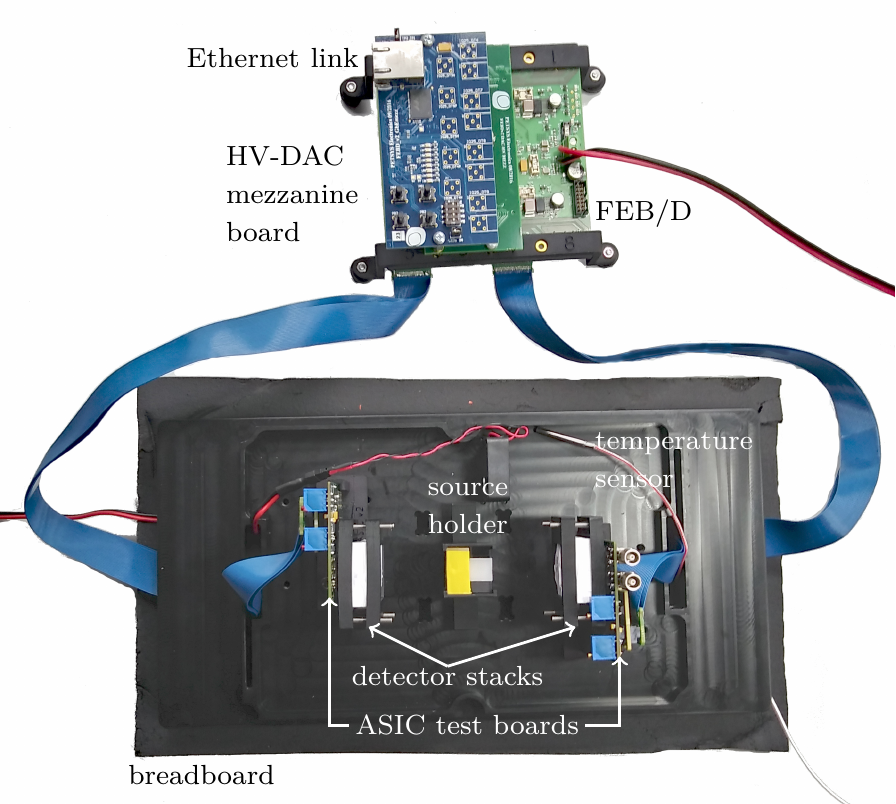}
	\caption{Benchtop setup included in the TOFPET2 ASIC evaluation kit showing two KETEK PA3325-WB-0808 SiPMs with 12-mm-high LYSO scintillator arrays connected to TOFPET2 ASIC evaluation boards set up for a coincidence experiment. The ASIC evaluation boards are connected to the FEB/D board which houses high-voltage regulators and a 1-Gbit-Ethernet communication interface.}
	\label{fig:setup}
\end{figure}

\subsection{Setup}
We used the TOFPET2 ASIC evaluation kit provided by PETsys Electronics S.A. (see Fig. \ref{fig:setup}) \cite{PETsysTOFPET2DataSheet,TOFPET2EKitHardwareGuide,TOFPET2EKitSoftwareGuide,PETsysFlyer}.  
The evaluation kit allows the user to test the TOFPET2 ASIC under benchtop conditions with different SiPM types.
The SiPMs can be connected to the ASIC via two SAMTEC connectors on the ASIC test boards shipped with the kit.
Apart from these test boards, the kit includes a front end board (FEB/D) holding the power supply and external clock for the ASIC test boards as well as a field programmable gate array (FPGA) and a 1-Gbit-Ethernet link for data transmission.
In addition, a high-voltage digital-to-analog converter (HV-DAC) mezzanine board is employed to provide the bias voltage for the SiPMs used.
The two ASIC test boards can be connected to the FEB/D board via two flexible cables and are mounted on a bread board for coincidence experiments (see Fig. \ref{fig:setup}).
The whole setup is enclosed by a light-impermeable top cover featuring a fan.
We added a temperature sensor connected to a controller allowing to adjust the ambient temperature of the setup.
The setup including FEB/D board is placed into a larger climate chamber, which is likewise thermally controlled.

\subsection{TOFPET2 ASIC}
\begin{figure}[t]
	\centering
	\includegraphics[height=0.5\columnwidth]{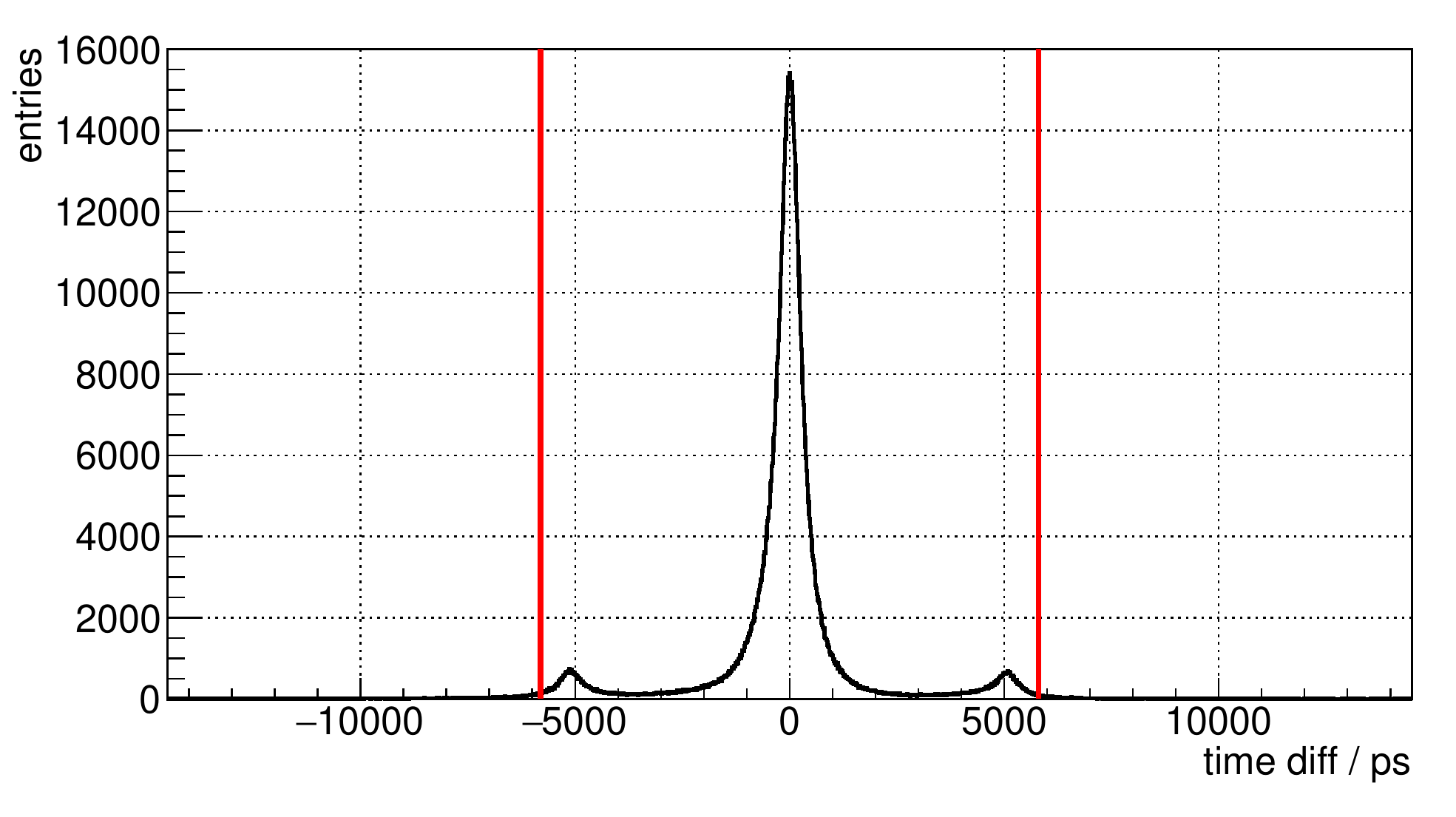}
	\caption{Time difference spectrum of two KETEK PA3325-WB-0808 SiPMs with 12-mm-high LYSO scintillator arrays in a coincidence experiment. Data were acquired at an overvoltage of \SI{4.75}{\volt} and with the discriminator thresholds set to $\mathsf{vth\_t1} = 10$, $\mathsf{vth\_t2} = 20$, and $\mathsf{vth\_e} = 15$. Red lines indicate the trigger delay period of \SI{5.9}{\nano\second}.}
	\label{fig:sat_peak_spec}
\end{figure}

The TOFPET2 ASIC (version 2b) features 64 individual channels, two TDCs with a time binning of \SI{30}{\pico\second}, and a clock cycle of \SI{200}{\mega\hertz} \cite{PETsysTOFPET2DataSheet}.
The user can choose a tot- or qdc-mode to measure the signal energy. 
The analog-to-digital converters (ADCs) used for the latter one are linear for integrating charges up to \SI{1500}{\pico\coulomb} (\SI{2500}{\photoelectron}) \cite{francesco2016TOFPET2}.
During acquisition, each channel is multi-buffered by four analog buffers.
The maximum event rate per ASIC channel is \SI{600}{\kilocountspersecond} \cite{PETsys_priv_comm}.
Each channel features an individual trigger circuit designed to reject dark counts by a three-threshold event validation.
Two discriminators $\mathsf{D\_T1}$ and $\mathsf{D\_T2}$ in the timing branch of the circuit are configured to trigger on different voltage thresholds, whereby the lower trigger of discriminator $\mathsf{D\_T1}$ is fed into an $\mathsf{AND}$ gate and validated by the higher trigger of discriminator $\mathsf{D\_T2}$.
A third discriminator $\mathsf{D\_E}$ with an even higher threshold is employed in the energy branch for further noise rejection.
Using the default trigger setting, an event is considered valid, if it triggers all three discriminators.
The voltage threshold of each discriminator can be adjusted via a dimensionless parameter $\mathsf{vth\_t1}$, $\mathsf{vth\_t2}$ or $\mathsf{vth\_e}$, respectively.
These parameters operate on different scales (approx. \SI{2.5}{\milli\volt}, \SI{15}{\milli\volt}, and \SI{20}{\milli\volt} per DAC step \cite{schug2018TOFPET2}) and adjust the voltage threshold over a channel-specific baseline determined in the calibration routine.
If an incident event only triggers $\mathsf{D\_T1}$, it is rejected.
On this first validation level, no dead time is introduced by the event rejection.
If an incident event also triggers  $\mathsf{D\_T2}$, the event timestamp is generated by this second trigger.
Hereafter, the event is validated or rejected by the third trigger.\\
It is possible that small pulses occurring just before coincidence events trigger $\mathsf{D\_T1}$ and are validated by the real event triggering $\mathsf{D\_T2}$ right after.
This causes the event timestamp to be generated by the output of $\mathsf{D\_T2}$ instead of the delayed output of $\mathsf{D\_T1}$ ($\mathsf{D\_T1'}$).
Subtracting two timestamps matched as a coincidence, where one was regularly assigned by $\mathsf{D\_T1'}$ and one was falsely generated by $\mathsf{D\_T2}$, results in a coincidence time difference modulated by the trigger delay period.
In the time difference spectra of matched coincidence events, these time differences are visible as satellite peaks shifted from the main peak by the trigger delay period (see Fig. \ref{fig:sat_peak_spec}).
A detailed description of the operation of the trigger circuit and the appearance of satellite peaks can be found in \cite{schug2018TOFPET2}.\\
Three software configuration parameters influence the input stage impedance $R_{\mathsf{IN}}$ and the discriminator noise of the TOFPET2 ASIC channel circuit, which both can be used to adjust the power consumption of the TOFPET2 ASIC.
All parameters operate on a dimensionless scale.
The parameters $\mathsf{fe\_ib1}$ and $\mathsf{fe\_ib2}$ affect the load on the signal line.
By changing $\mathsf{fe\_ib1}$, the input impedance can be adjusted.
The impedance is exponentially increased from \SIrange{11}{32}{\ohm} when $\mathsf{fe\_ib1}$ is changed from 0 to 60, which reduces a current $\mathsf{I_{B1}}$ in the preamplifier circuit  \cite{PETsysTOFPET2DataSheet}.  
Considering a parasitic capacitance on this line, a higher input stage impedance leads to a slower signal.
The discriminator noise $\mathsf{V_{noise\_T}}$ and the discriminator noise slew rate, respectively, can be adjusted by changing the parameters $\mathsf{fe\_ib2}$ and $\mathsf{disc\_sf\_bias}$.
The parameter $\mathsf{fe\_ib2}$ modifies a current $\mathsf{I_{B2}}$ in the preamplifier circuit, which changes the signal amplification.
Details on the preamplifier circuit, which itself consumes \SI{2.5}{\mWperchannel}, are given in \cite{francesco2016TOFPET2}.
The parameter $\mathsf{disc\_sf\_bias}$ affects the biasing of signal buffers between two blocks, in which the discriminators are divided.
Reducing the buffer biasing by choosing a smaller value for $\mathsf{disc\_sf\_bias}$, we expect a slower internal copy of the signal.
We use the nomenclature from the PETsys documentation \cite{PETsysTOFPET2DataSheet}.
No numerical values are given regarding the impact of $\mathsf{fe\_ib2}$ and $\mathsf{disc\_sf\_bias}$ on physical parameters \cite{PETsysTOFPET2DataSheet}.

\subsection{$\gamma$-Detectors}
In this study, three configurations of SiPMs and scintillators were used with the setup.
For single-channel experiments, we employed two FBK (NUV-HD) SiPMs each coupled to a \SI{2.62x2.62x3}{\mm} LYSO scintillation crystal using Cargille Meltmount\texttrademark~($\mathsf{n_D} = 1.539$).
To connect these to the ASIC test boards, two small adapter boards provided by PETsys were used.
A \textsuperscript{22}Na point source (\SI{0.5}{\mm} active diameter) with an activity of approx. \SI{7}{\mega\becquerel} was placed in the center of the setup.
For multi-channel experiments, we employed two 8 $\times$ 8 KETEK PA3325-WB-0808 SiPM arrays or two 8 $\times$ 8 Hamamatsu S14161-3050-HS-08, each one-to-one coupled to a 12-mm-high scintillator array, featuring BaSO$_4$ powder mixed with epoxy as the inter-crystal layer. %
An individual crystal has the dimensions \SI{3x3x12}{\milli\m}.
For optical coupling, Sylgard\textsuperscript{\textregistered} 527, a two-component dielectric gel fabricated by Dow Corning, was used.
A geometry of five \textsuperscript{22}Na NEMA cubes (\SI{0.25}{\mm} active diameter, \SI{10}{\mm} edge length, suggested to be used for resolution testing according to the NEMA NU4-2008 standards \cite{NEMANU4resolutiontesting}) with a total activity of approx. \SI{3}{\mega\becquerel} was placed in the center of the setup.
All single- and multi-channel configurations were wrapped in teflon tape to prevent light loss.

\section{Methods}

In this work, the power consumption of the TOFPET2 ASIC was quantified, including a study of the range of possible configurations.
The power consumption of the TOFPET2 ASIC is a function of the input stage impedance, the discriminator noise, and its slew rate.
Adjustments of the configuration parameters are expected to affect the ASIC performance.
This performance impact was also evaluated.
The available configuration parameters do not allow to adjust the input capacitance, which influences the ASIC power consumption as well, but cannot be configured \cite{PETsysTOFPET2DataSheet}.

\subsection{Experiments}

\begin{figure}[t]
	\centering
	\includegraphics[width=\columnwidth]{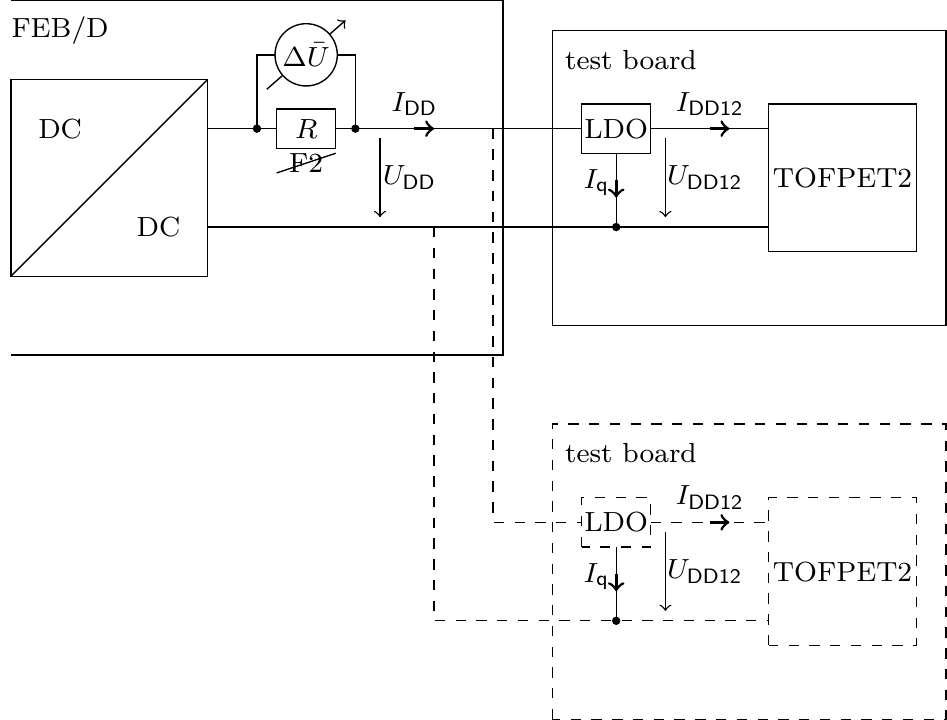}
	\caption{Schematic drawing of the circuit used for the power consumption measurement. The voltage drop $\Delta \bar{U}$ in the 1.2-V-line is measured over a small resistance $R$ added to the power supply line. The bias supply as well as the 2.5-V-line are not shown. Optionally, a second ASIC test board can be connected to the line.}
	\label{fig:schematic_circuit}
\end{figure}

The ASIC requires a power supply of \SI{1.2}{\volt} powering the ASIC operation and a power supply of \SI{2.5}{\volt} powering the ASIC-FPGA communication \cite{PETsysTOFPET2DataSheet}.
The current consumption of the ASIC on the 1.2-V-line is not conveniently measurable, since this voltage (\SI{1.2}{\volt}) is generated locally on the ASIC test board (see Fig. \ref{fig:schematic_circuit}). 
Here, a linear low-dropout (LDO) regulator with a negligible quiescent current $I_q$ is used and the internal circuits of that LDO are fed out of a different supply line.
Hence, the average of the ASIC supply current $I_{\mathsf{DD12}}$ is basically identical to the average value of LDO input current $I_{\mathsf{DD}}$. 
This input current can be measured via a modification of the FED/D board. 
The FEB/D generates a pre-regulated voltage $U_{\mathsf{DD}}$ of \SI{1.8}{\volt}. 
This voltage is protected by a replaceable fuse F2. 
Exchanging this fuse with a shunt resistor $R$ allows to evaluate the current via measuring the voltage drop across this resistor. 
For our test, we used a shunt resistor of \SI{0.13}{\ohm} and captured the voltage drop $ \Delta \bar{U}$ with an oscilloscope. 
Mathematical averaging reveals the average current. 
The output voltage of the LDO was confirmed to stay unaffected from the modification, i.e., the voltage drop across $R$ did not impact the output voltage regulation, which was stable at \SI{1.2}{\volt}. 
Hence, the power consumption $\bar{P}$ of the ASIC was then computed via
\begin{equation}
	\bar{P} = U_{\mathsf{DD12}} \cdot \bar{I}_{\mathsf{DD12}}  = U_{\mathsf{DD12}} \cdot \frac{\Delta \bar{U}}{R} 
\end{equation}
in case a single ASIC board was connected to that very FEB/D channel. 
As visible in Fig. \ref{fig:schematic_circuit}, a second ASIC board can be connected to the same output. 
In this case, the calculated power is hence the power for two ASICs.
The introduced method only measures the ASIC power consumption due to operating the ASIC itself.
The power consumption due to ASIC-FPGA communication is not included in the measured values, since it cannot be separated from other loads on the 2.5-V-line, e.g., the FPGA.\\
The three software configuration parameters $\mathsf{fe\_ib1}$, $\mathsf{fe\_ib2}$, and $\mathsf{disc\_sf\_bias}$ were successively changed to evaluate their impact on the power consumption.
So far, only $I_\mathsf{DD12}$ as a function of $\mathsf{fe\_ib2}$ and $\mathsf{disc\_sf\_bias}$ is reported \cite{PETsysTOFPET2DataSheet}.
This study intends to provide a complete overview on the impact of the configuration parameters.
The parameter $\mathsf{fe\_ib1}$ was changed from 0 to 60 in steps of \SI{5}{}. 
The parameter $\mathsf{fe\_ib2}$ was changed from 0 to 30 in steps of \SI{5}{}.
The parameter $\mathsf{disc\_sf\_bias}$ was changed from 0 to 32 in steps of \SI{4}{}.
While one parameter was changed, the other two are kept at zero to reveal the influence of a that very parameter.
For each setting, the respective power consumption was determined for data acquisition running in tot- or qdc-mode in single- and multi-channel experiments with FBK (NUV-HD) and KETEK PA3325-WB-0808 SiPMs and for acquiring dark counts or events of \textsuperscript{22}Na sources placed inside the setup.
For all applied settings, data were acquired for \SI{30}{\second} at \SI{16}{\celsius} ambient temperature.
The overvoltage was set to \SI{4.75}{\volt}.
The discriminator thresholds were kept constant at $\mathsf{vth\_t1} = 20$, $\mathsf{vth\_t2} = 20$, and $\mathsf{vth\_e} = 15$.
The minimum, default, and maximum power consumption of the ASIC was determined as benchmarks for further investigations. 
The corresponding parameter settings are $\mathsf{fe\_ib1} = 60$, $\mathsf{fe\_ib2} = 30$, and $\mathsf{disc\_sf\_bias} = 32$ to reach minimum, $\mathsf{fe\_ib1} = 59$, $\mathsf{fe\_ib2} = 0$, and $\mathsf{disc\_sf\_bias} = 0$ to reach default, and $\mathsf{fe\_ib1} = 0$, $\mathsf{fe\_ib2} = 0$, and $\mathsf{disc\_sf\_bias} = 0$ to reach maximum power consumption.
The settings for minimum and maximum power consumption were determined experimentally.
The setting for default power consumption was extracted from the default ASIC software configuration.\\
Additionally, the impact of the power consumption setting on the ASIC performance was evaluated in coincidence experiments with two KETEK PA3325-WB-0808 SiPM arrays.
For minimum, default, and maximum power consumption, the overvoltage was varied between \SIrange{2.75}{5.75}{\volt} in steps of \SI{0.5}{\volt} with $\mathsf{vth\_t1} = 30$, $\mathsf{vth\_t2} = 20$, and $\mathsf{vth\_e} = 15$.
For each setting, data were acquired for \SI{120}{\second}.\\
Employing Hamamatsu S14161-3050-HS-08 SiPM arrays, the influence on the dark count rate per channel was investigated.
For this purpose, each channel was individually enabled to trigger only on the first discriminator $\mathsf{D\_T1}$ of the ASIC channel circuit.
The validation by higher thresholds was disabled.
For each setting, dark counts were acquired for \SI{10}{\second} at an overvoltage of \SI{4.75}{\volt} and for discriminator thresholds between $\mathsf{vth\_t1} = 1$ and $\mathsf{vth\_t1} = 60$ in steps of 1.

\subsection{Setup calibration}
A calibration according to the PETsys calibration routine \cite{TOFPET2EKitSoftwareGuide} was run for each investigated SiPM type at default ASIC configuration with an overvoltage of \SI{4}{\volt} and  $\mathsf{vth\_t1} = 20$, $\mathsf{vth\_t2} = 20$, and $\mathsf{vth\_e} = 15$.
Investigating the impact of the power consumption on the ASIC performance and the dark count rate, a calibration was run at each power consumption setting.\\

\subsection{Data processing}
Performance data were evaluated in the same manner as described in \cite{schug2018TOFPET2}: 
Data were prepared applying the PETsys routine $\mathsf{convert\_raw\_to\_singles}$ to convert raw data into single hit information.
The obtained table containing a timestamp, an energy value, and a channel ID for each single hit registered was used for further processing.
An energy value histogram was computed, where the peak positions of the two peaks in the \textsuperscript{22}Na spectrum (\SI{511}{\keV}, \SI{1274.5}{\keV}) were determined using a Gaussian fit routine.
A saturation corrected model was fit to the determined positions, allowing to compute the energy in keV from the acquired energy values via
\begin{equation}
	E = c \cdot s \cdot \log \left( \frac{1}{1 - \frac{e}{s}} \right)
\end{equation}
Here, $E$ is the hit energy in keV and $e$ is the energy value in ADC units acquired for this hit. 
The factors $c$ and $s$ are a conversion factor and a saturation parameter determined by the fit routine.
The energy resolution was determined as the full width at half maximum (FWHM) of the \SI{511}{\keV} peak in the converted energy spectrum.
Single hits were checked for coincidences applying an energy filter of \SIrange{400}{700}{\keV} and a coincidence window of \SI{7.5}{\nano\second}. 
For multi-channel data, the timestamps were corrected for the source positions.
The coincidence time difference between two matched hits was computed.
From the time difference histogram, the coincidence resolution time (CRT) was computed as the FWHM of the histogram peak.
For each time difference histogram, the satellite peak fraction is computed.
This fraction classifies all events matched as coincidences with a coincidence time difference larger than \SI{2.5}{\nano\second}.\\
Performance results are stated dependent on the relative offset-corrected overvoltage \ac{Ucorrel}, which is computed via 
\begin{equation}
U_{\mathsf{cor,rel}} = \frac{U_{\mathsf{bias,set}} - U_{\mathsf{off}} - \bar{U}_{\mathsf{BD}} }{\bar{U}_{\mathsf{BD}}},
\end{equation}
where \ac{Ubd} is the breakdown voltage of the employed SiPM, \ac{Ubias} is the applied bias voltage configured via the software, and \ac{Uoff} is the voltage offset between configured and actually applied bias that was determined by probing different ASIC channels. The voltage offset is due to a small DC voltage (approx. \SI{750}{\milli\volt}) at the input of each ASIC channel \cite{TOFPET2EKitSoftwareGuide2019}.

\section{Results}

\subsection{Adjustability}

\begin{figure}[t]
	
	\centering
	\includegraphics[width=0.95\columnwidth]{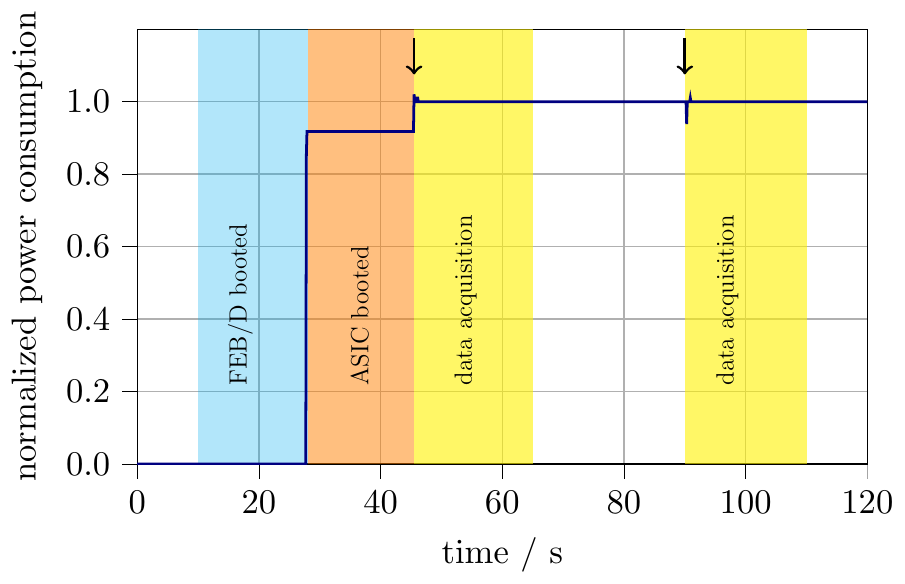}
	\caption{Normalized power consumption on the 1.2-V line accumulated for 128 ASIC channels when switching between different system states at default configuration of the ASIC power consumption. Data were acquired with two KETEK PM3325-WB-0808 and no radioactive sources inside the setup.}
	\label{fig:power_consumption_state_switching}
\end{figure}

\begin{figure*}[t]
	\centering
	\tikzsubfig{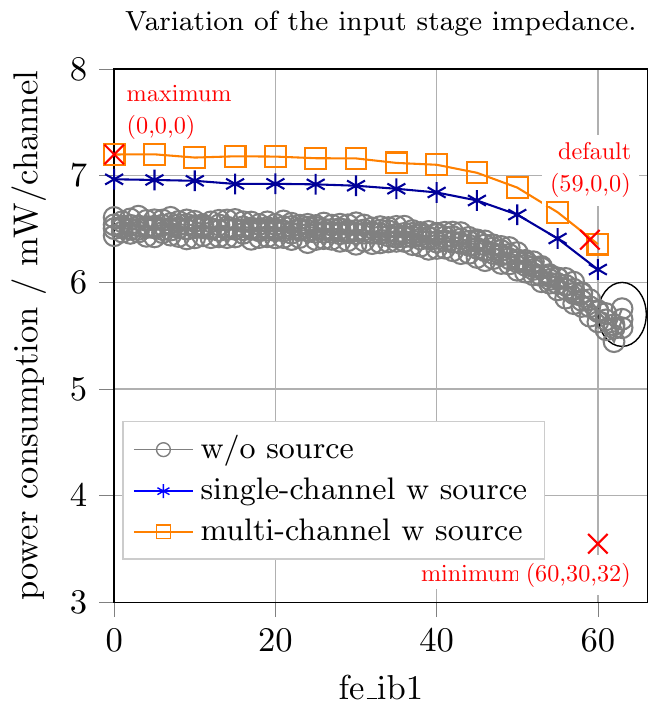}{width=0.65\columnwidth}{fig:fig05a}
	\tikzsubfig{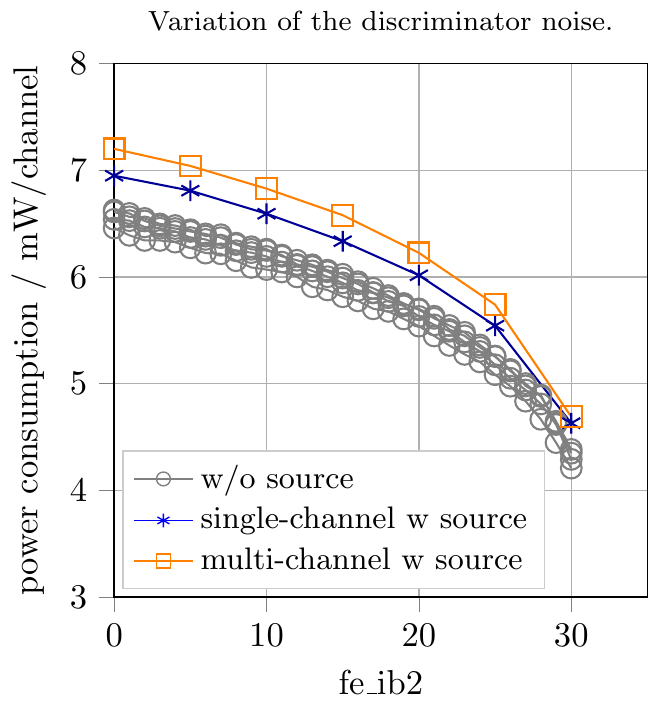}{width=0.65\columnwidth}{fig:fig05b}
	\tikzsubfig{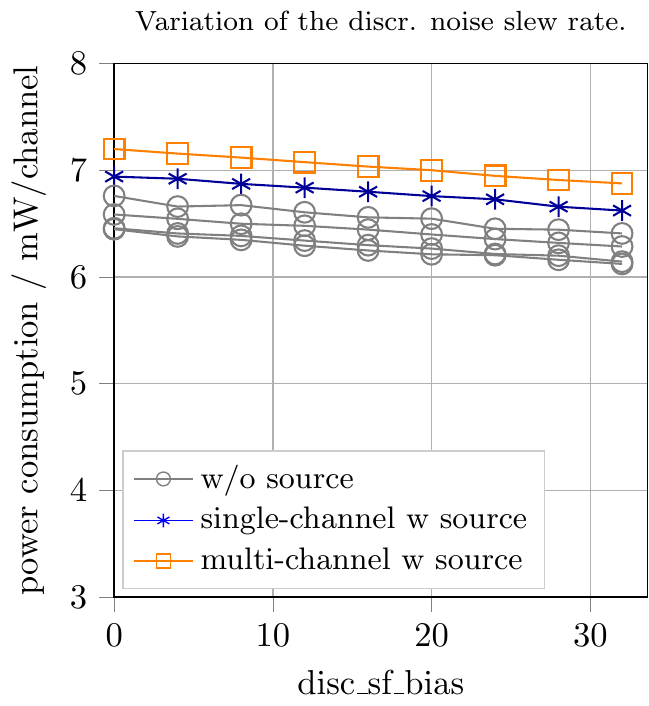}{width=0.65\columnwidth}{fig:fig05c}
	
	\caption{\protect\subref{fig:fig05a} Variation of input stage impedance.
		\protect\subref{fig:fig05b} Variation of the discriminator noise.
		\protect\subref{fig:fig05c} Variation of the discriminator noise slew rate.
		Power consumption in qdc-mode. Data are acquired in single- (FBK (NUV-HD) SiPMs) and multi-channel (KETEK PA3325-WB-0808) experiments. The measured values exclude the power consumption due to FPGA-ASIC communication. The relative statistical error on the measured power consumption is approx. \SI{8}{\percent}. The black circle in \protect\subref{fig:fig05a} indicates a small increase in power consumption for $\mathsf{fe\_ib1} > 60$ ($R_{\mathsf{IN}} > \SI{32}{\ohm}$).}
	\label{fig:pow_consump_qdc}
\end{figure*}

The power consumption on the 1.2-V line changes for switching between system states (whole setup turned off, FEB/D booted, ASICs booted, measurement running, see Fig. \ref{fig:power_consumption_state_switching}).
It stays constant during data acquisition, i.e., for the channel trigger circuit switching between different trigger states, and also between multiple measurements.
Peaks visible when transferring the ASIC configuration for a new measurement (indicated by arrows in Fig. \ref{fig:power_consumption_state_switching}) are small compared to the total power consumption (approx. \SI{6}{\percent} change).\\
In Fig. \ref{fig:pow_consump_qdc}, we depict the influence of each of the three software configuration parameters $\mathsf{fe\_ib1}$, $\mathsf{fe\_ib2}$, and $\mathsf{disc\_sf\_bias}$ on the TOFPET2 ASIC  power consumption in qdc-mode.
In tot-mode, the acquired curves show the same shape.
All parameters cause a drop of the power consumption when being increased.
Here, the software configuration parameter $\mathsf{fe\_ib2}$, which, according to the data sheet \cite{PETsysTOFPET2DataSheet}, influences the discriminator noise, has the largest impact on the power consumption.
The power consumption drops from \SIrange{6.5}{4.3}{\mWperchannel} for the acquisition of dark counts.
The parameters $\mathsf{fe\_ib1}$ and $\mathsf{disc\_sf\_bias}$ have a lower impact on the power consumption.
Here, the power consumption drops from \SIrange{6.5}{5.8}{\mWperchannel} and from \SIrange{6.5}{6.2}{\mWperchannel}, respectively, for the acquisition of dark counts.
No difference is visible between single- and multi-channel experiments.
In these scans, we notice a systematic increase of the power consumption when configuring $\mathsf{fe\_ib1} > 60$, i.e., $R_{\mathsf{IN}} > \SI{32}{\ohm}$ (see black circle in Fig. \ref{fig:fig05a}).
As a consequence, settings with $\mathsf{fe\_ib1}> 60$ are excluded from further scans.
If events of a radioactive source are acquired, a slight but systematic increase of the power consumption of approx. \SI{0.5}{\mWperchannel} is visible.
A shift of again approx. \SI{0.2}{\mWperchannel} is visible if an SiPM array is connected to the ASIC test board instead of a single SiPM.
Statistical errors of approx. \SI{8}{\percent} mainly stemming from the resistance measurement are equally assumed on all measured data.
As benchmarks for further investigations and for comparison with other ASIC models, the power consumption in qdc-mode is determined to be \SI{3.6}{\mWperchannel} at its minimum, \SI{6.4}{\mWperchannel} at its default, and \SI{7.2}{\mWperchannel} at its maximum value (see Tab. \ref{tab:benchmarks}).

\subsection{Analytical model}

\begin{figure}[t]
	\centering
	\includegraphics[width=0.9\columnwidth]{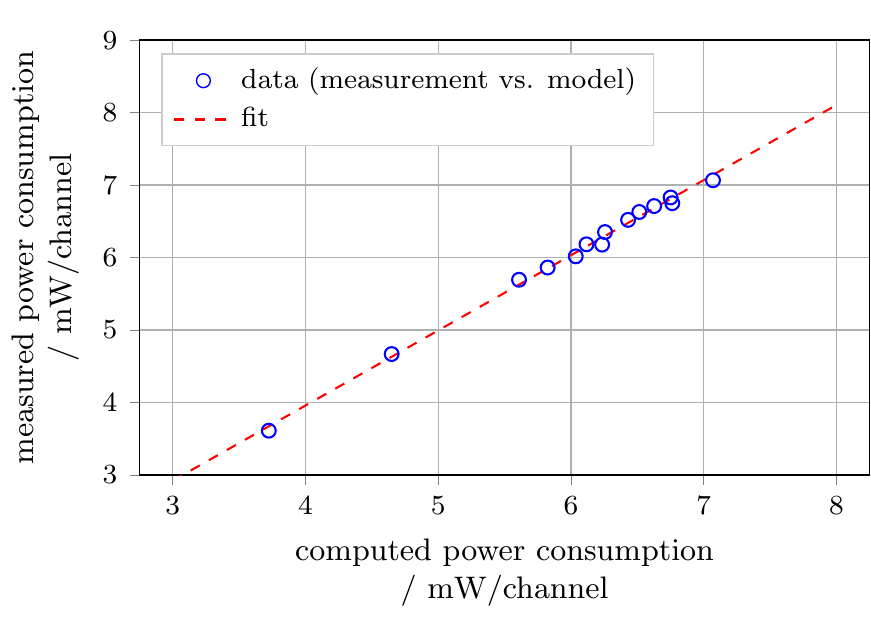}
	\caption{Power consumption at various randomly chosen settings of $\mathsf{fe\_ib1}$, $\mathsf{fe\_ib2}$ and $\mathsf{disc\_sf\_bias}$. The measured power consumption is plotted against the power consumption computed by applying the introduced analytical model. The linearity of the fit (red line) is determined to be 1.034 $\pm$ 0.018. Data are acquired with two KETEK PA3325-WB-0808 SiPM arrays.}
	\label{fig:fig06}
\end{figure}

We assume that the power consumption per channel $P_{\mathsf{ch}}$ can be computed analytically prior to experimental determination by applying a model linearly superposing the observed effects for a given parameter tuple ($\mathsf{fe\_ib1}$, $\mathsf{fe\_ib2}$,  $\mathsf{disc\_sf\_bias}$).
To determine the model parameters, piece-wise defined functions are fit to the curves.
A constant \ac{P0} is assumed that is set off against the allocated influence of the three parameters $\mathsf{fe\_ib1}$, written as \ac{dPfeib1}, $\mathsf{fe\_ib2}$, written as \ac{dPfeib2}, and $\mathsf{disc\_sf\_bias}$, written as \ac{dPdiscsfbias}:
\begin{equation}
P_{\mathsf{ch}} = P_0  + dP_{\mathsf{fe\_ib1}} + dP_{\mathsf{fe\_ib2}} + dP_{\mathsf{disc\_sf\_bias}}
\end{equation}
We use linear or parabolic functions to model the impact of the respective parameters. 
When changing $\mathsf{disc\_sf\_bias}$ over its whole parameter range, the power consumption drops linearly (see Fig. \ref{fig:fig05c}).
Hence, \ac{dPdiscsfbias} can be parameterized as
\begin{equation}
dP_{\mathsf{disc\_sf\_bias}} =
\begin{cases} 
 f_0 \cdot \mathsf{disc\_sf\_bias},& \\
\quad 0 \leq \mathsf{disc\_sf\_bias} \leq 32 &\\
\end{cases}
\end{equation}
For $\mathsf{fe\_ib1}$ and $\mathsf{fe\_ib2}$, the drop in power consumption is linear first, but becomes parabolic towards higher parameter values (see Fig. \ref{fig:fig05a} and Fig. \ref{fig:fig05b}).
Therefore, the influence of $\mathsf{fe\_ib1}$ and $\mathsf{fe\_ib2}$ is parameterized as:
\begin{equation}
dP_{\mathsf{fe\_ib1}} =
\begin{cases} 
 a_1 \cdot \mathsf{fe\_ib1}, &  0 \leq \mathsf{fe\_ib1} \leq 40\\
 & \\
 a_2 \cdot \mathsf{fe\_ib1}^2 & \\
\quad  + a_3 \cdot \mathsf{fe\_ib1}, & 41 \leq \mathsf{fe\_ib1} \leq 60 \\
\end{cases}
\end{equation}
\begin{equation}
dP_{\mathsf{fe\_ib2}} =
\begin{cases} 
 b_0 \cdot \mathsf{fe\_ib2}, &  0 \leq \mathsf{fe\_ib2} \leq 20\\
 & \\
 b_1 \cdot \mathsf{fe\_ib2}^2  &  \\
\quad + b_2 \cdot \mathsf{fe\_ib2}, & 21 \leq \mathsf{fe\_ib2} \leq 30 \\
\end{cases}
\end{equation}
The model parameters $a_i$, $b_i$, and $f_0$ are determined using least-squares fit routine (see Tab. \ref{tab:parameters_linear_model}).
As $P_0$, i.e., the y-axis intercept of the power consumption curves in Fig. \ref{fig:pow_consump_qdc}, shows a dependency on the count rate, it has to be determined experimentally setting $\mathsf{fe\_ib1} = 0$, $\mathsf{fe\_ib2} = 0$, and $\mathsf{disc\_sf\_bias} = 0$.\\
In order to test this model, random tuples ($\mathsf{fe\_ib1}$, $\mathsf{fe\_ib2}$,  $\mathsf{disc\_sf\_bias}$) are considered.
The power consumption of the ASIC is measured for each tuple using the methods applied before and computed via the implemented model.
The measured power consumption is plotted against the computed power consumption (see Fig. \ref{fig:fig06}, blue dots).
A linear regression is performed on the data points (see Fig. \ref{fig:fig06}, red line).
The linearity of the fit is determined to be 1.034 $\pm$ 0.018 with an y-axis intercept in the order of \SI{0.1}{\mWperchannel}.

\begin{table}[t]
	\caption{Power consumption of the TOFPET2 ASIC measured at minimum, default, and maximum setting. The parameters $\mathsf{fe\_ib1}$, $\mathsf{fe\_ib2}$, and $\mathsf{disc\_sf\_bias}$ are software configuration variables.}
	\label{tab:benchmarks}
	
	\centering
	\begin{tabular}{|l|c|c|c|c|}
		\hline
		& $\mathsf{fe\_ib1}$ & $\mathsf{fe\_ib2}$ &  $\mathsf{disc\_sf\_bias}$ & $P$ / mW/channel \\
		\hline
		minimum & 60 & 30 & 32 & 3.6\\
		default & 59 & 0 & 0 & 6.4\\
		maximum & 0 & 0 & 0 & 7.2\\
		\hline
	\end{tabular}
	
\end{table}

\begin{table}[t]
	
	\caption{Parameters of the analytical model to compute the TOFPET2 power consumption from software configuration parameters. Parameters are computed mean values of the fit parameters determined using a least-squares fit routine.}
	\label{tab:parameters_linear_model}
	\centering
	\begin{tabular}{|p{2cm}|r|}
		\hline
		parameter & value /\SI{}{\mWperchannel} \\
		\hline
		\ac{f0} & $(-9.96 \pm 0.03) \cdot 10^{-3}$\\
		\ac{a0} & $(-2.53 \pm 0.02) \cdot 10^{-3}$\\
		\ac{a1} & $(-0.60 \pm 0.01) \cdot 10^{-3}$\\
		\ac{a2} & $(23.17 \pm 1.12) \cdot 10^{-3}$\\
		\ac{b0} & $(-45.41 \pm 0.17) \cdot 10^{-3}$\\
		\ac{b1} & $(2.74 \pm 0.02) \cdot 10^{-3}$\\
		\ac{b2} & $(6.70 \pm 0.86) \cdot 10^{-3}$\\\hline \hline
		$P_0^{*}$ & 7.07\\
		\hline
		\multicolumn{2}{l}{\scriptsize $^{*}$Value was determined experimentally.}
	\end{tabular}
	
\end{table}

\subsection{Stability}
\begin{figure}[!t]
	\centering
	\includegraphics[width=\columnwidth]{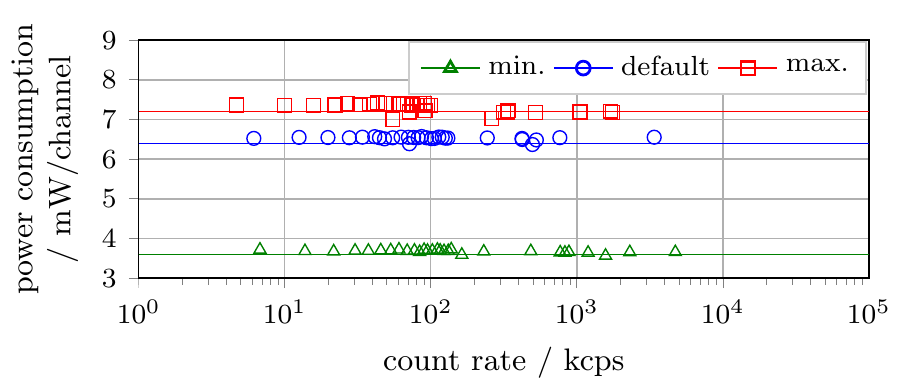}
	\caption{Stability of the TOFPET2 ASIC power consumption for changing count rates at the three benchmarks determined in previous measurements. The stated count rates refer to the total number of validated events at setup level. Count rate variations were achieved by enabling different numbers of ASIC channels to trigger, applying different discriminator thresholds $\mathsf{vth\_t1}$ or varying the distance to the source. All measurements were conducted with two KETEK PA3325-WB-0808 SiPM arrays.}
	\label{fig:fig07}
\end{figure}

\begin{figure}[!t]
	\centering
	\includegraphics[width=\columnwidth]{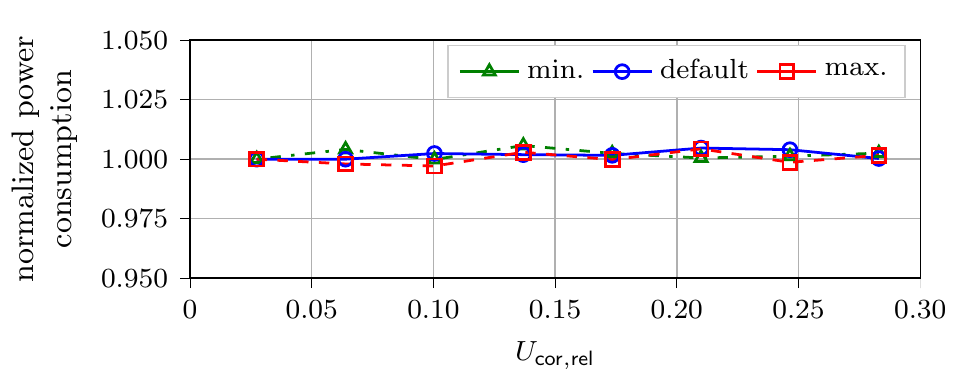}
	\caption{Stability of the overvoltage at the three benchmarks of the TOFPET2 ASIC power consumption determined in previous measurements. The power consumption is normalized to the value acquired for an overvoltage of \SI{0.75}{\volt} at each benchmark. All measurements were conducted with one KETEK PA3325-WB-0808 SiPM array. No sources were places inside the setup.}
	\label{fig:fig08}
\end{figure}

The power consumption per channel is shown to be stable for overvoltages ranging from \SI{0.75}{\volt} to \SI{7.75}{\volt} at the three benchmark settings determined in prior measurements (see Fig. \ref{fig:fig08}).
In addition, the power consumption is stable for various count rates at the three benchmark settings (see Fig. \ref{fig:fig07}).
Neither different discriminator thresholds $\mathsf{vth\_t1}$ and different source distances nor changing numbers of ASIC channels enabled to trigger were observed to change the measured power consumption per channel significantly.

\subsection{Impact on performance}

\begin{figure}[!t]
	\centering
	\includegraphics[width=0.9\columnwidth]{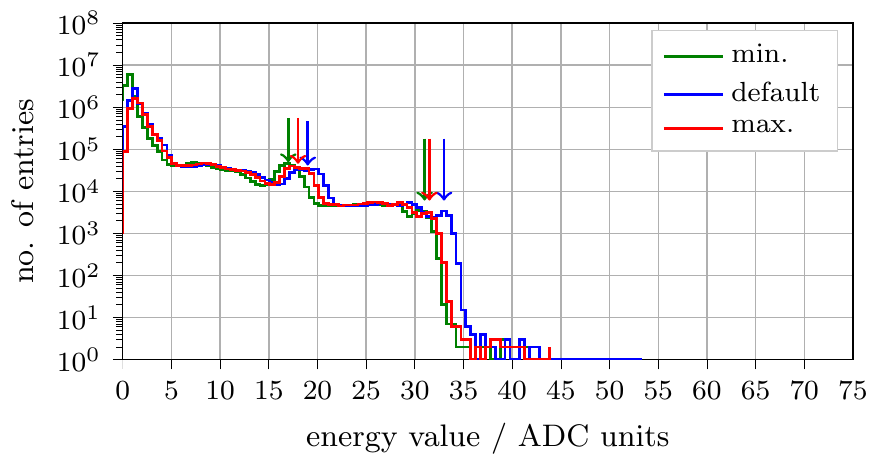}
	\caption{Energy spectra at the three benchmarks: \textcolor{black}{minimum} (\SI{3.6}{\mWperchannel}), \textcolor{black}{default} (\SI{6.4}{\mWperchannel}) and \textcolor{black}{maximum} (\SI{7.2}{\mWperchannel}). Energy spectra were acquired with two KETEK PA3325-WB-0808 SiPM arrays at \SI{4.75}{\volt} overvoltage and with $\mathsf{vth\_t1}=50$. Arrows indicate the positions of the \SI{511}{\keV} and \SI{1274.5}{\keV} peaks in the energy value spectra.}
	\label{fig:power_consumption_vs_energy_spectrum}
\end{figure}

\begin{figure}[!t]
	\centering
	\tikzsubfig{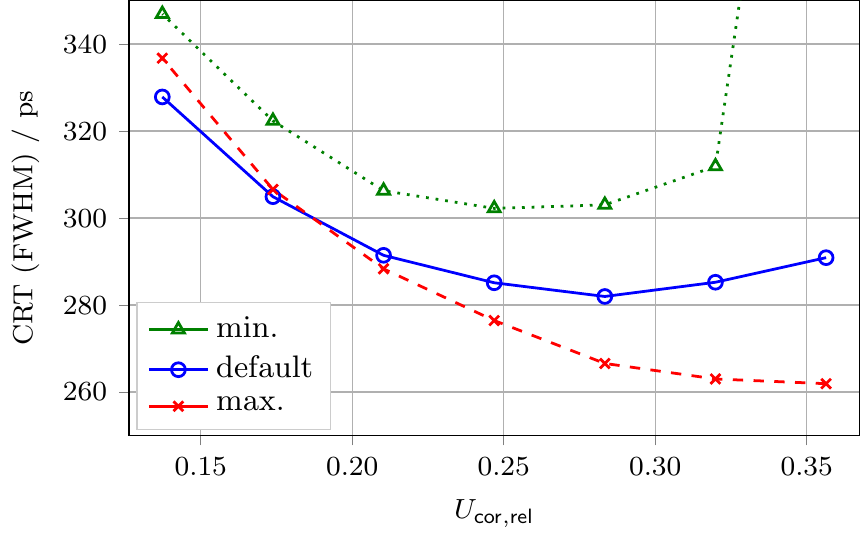}{width=0.8\columnwidth}{fig:pow_consump_vs_performance_CRT}
	\\
	\tikzsubfig{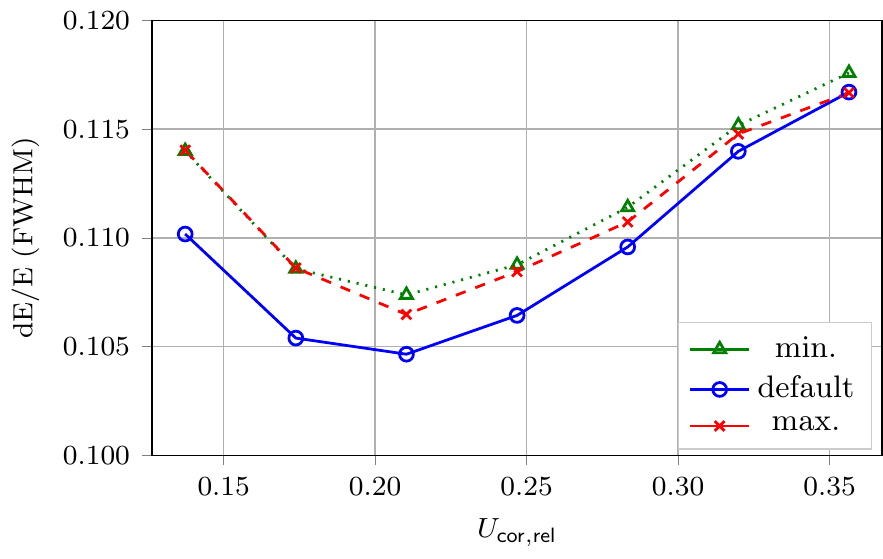}{width=0.8\columnwidth}{fig:pow_consump_vs_performance_eres}
	\\
	\tikzsubfig{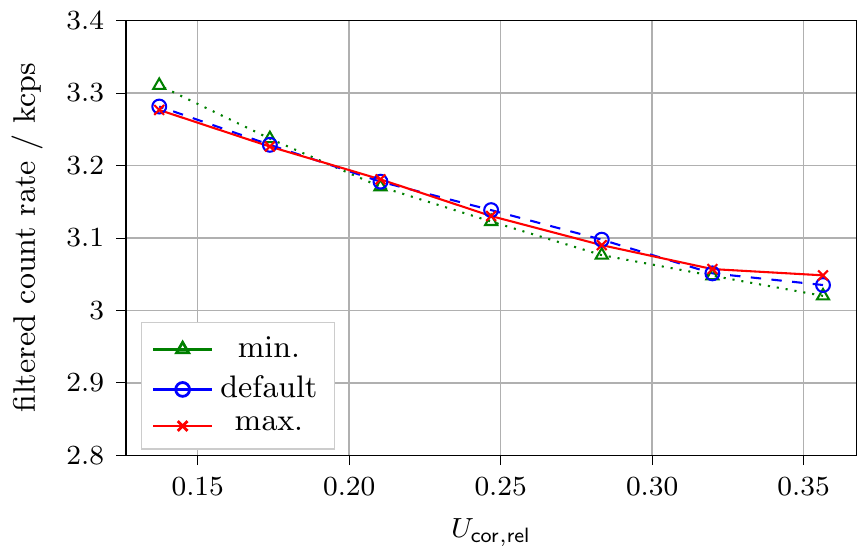}{width=0.8\columnwidth}{fig:pow_consump_vs_performance_filteredcps}
	\\
	\tikzsubfig{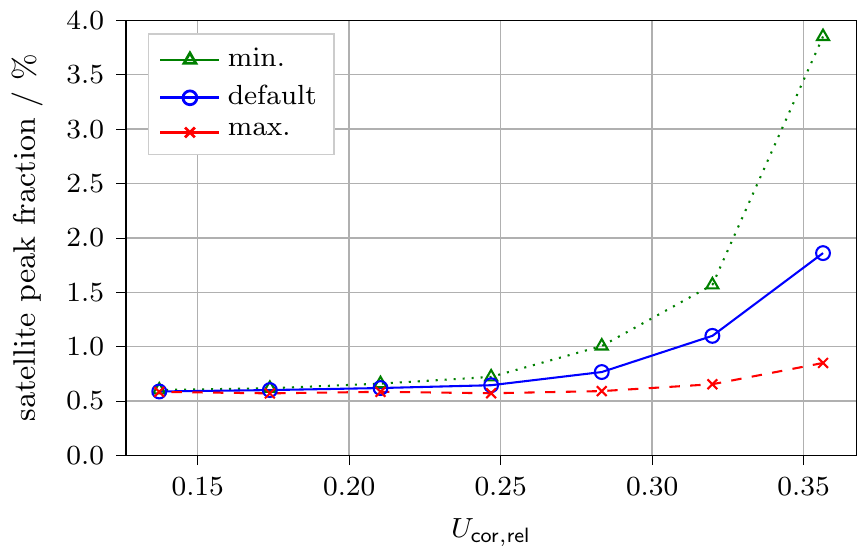}{width=0.8\columnwidth}{fig:fig10ds}
	\caption{\protect\subref{fig:pow_consump_vs_performance_CRT} Coincidence resolution time (CRT).
		\protect\subref{fig:pow_consump_vs_performance_eres} Energy resolution (dE/E).
		\protect\subref{fig:pow_consump_vs_performance_filteredcps} Filtered count rate.
		\protect\subref{fig:fig10ds} Satellite peak fraction.
		Power consumption vs. performance at the three benchmarks: \textcolor{black}{minimum} (\SI{3.6}{\mWperchannel}), \textcolor{black}{default} (\SI{6.4}{\mWperchannel}) and \textcolor{black}{maximum} (\SI{7.2}{\mWperchannel}). Measurements were performed with two KETEK PA3325-WB-0808 SiPM arrays at \SI{16}{\celsius} and for $\mathsf{vth\_t1}=30$, $\mathsf{vth\_t2}=20$, and $\mathsf{vth\_e}=15$ in qdc-mode. Statistical errors of up to \SI{1}{\pico\second} on the CRTs, up to \SI{0.02}{\percent} (absolute error) on the energy resolution, and up to \SI{0.03}{\percent} (absolute error) on the satellite peak fraction are reported.}
	\label{fig:pow_consump_vs_performance}
\end{figure}

\begin{figure}[t]
	\centering
	\includegraphics[width=0.92\columnwidth]{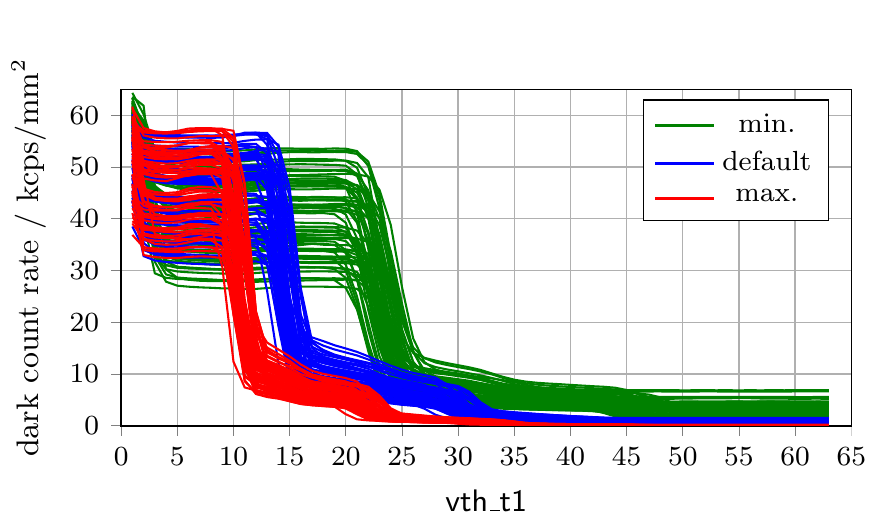}
	\caption{Dark count rate as a function of the first discriminator threshold of a Hamamatsu S14161-3050-HS-08 SiPM array coupled to a 12-mm-high LYSO scintillator array featuring \SI{360}{\micro\metre} BaSO$_4$ power mixed with epoxy as the inter-crystal layer. Different curves of the same color indicate different ASIC channels. Data were acquired at an overvoltage of \SI{4.75}{\volt} acquired in qdc-mode. The ambient temperature is set to \SI{16}{\celsius}. The power consumption is configured to be at its minimum, default and maximum value (see Tab. \ref{tab:benchmarks} for settings and values).}
	\label{fig:fig011}
\end{figure}

The position of the \SI{511}{\keV} and  \SI{1274.5}{\keV} peaks of the energy value spectra acquired via signal integration (qdc-mode) is affected by different power consumption settings as indicated by arrows in Fig. \ref{fig:power_consumption_vs_energy_spectrum}.
The filtered count rate (counts with an energy between \SIrange{400}{700}{\keV}) drops by approx. \SI{10}{\percent} over the overvoltage range investigated (see Fig. \ref{fig:pow_consump_vs_performance_filteredcps}). 
The acquired filtered count rate does not change for different power consumption settings.
Furthermore, a slight deterioration in energy resolution is observed in performance experiments (see Fig. \ref{fig:pow_consump_vs_performance_eres}).
In comparison to the default setting, systematic deviations smaller than \SI{0.5}{\percent} (absolute change) are visible for the energy resolution at maximum and minimum power consumption for all applied overvoltages.\\
Adjusting the power consumption is shown to have a significant influence on the CRTs achieved in coincidence experiments (see Fig. \ref{fig:pow_consump_vs_performance_CRT}).
At the cost of a higher consumption, CRTs can be improved by \SIrange{20}{40}{\pico\second} comparing the performance for different overvoltages at minimum and maximum power consumption configuration.\\
Additionally, a higher fraction of events contributing to the formation of satellite peaks in the coincidence time difference spectra is reported for a lower power consumption. 
The fraction increases by up to \SI{3}{\percent} comparing the situation for minimum and maximum power consumption at different overvoltages (see Fig. \ref{fig:fig10ds}).
Since prior studies showed that the satellite peak fraction depends on the configured trigger threshold of the first discriminator \cite{schug2018TOFPET2}, the  increased fraction calls for an adjustment of the trigger thresholds during performance experiments.\\
Additionally, the dark count rate was acquired for settings in the full range of possible thresholds $\mathsf{vth\_t1}$. 
Scanning the dark counts of an SiPM in qdc-mode results in acquiring the number of events, i.e., the number of SiPM pulses at and above a configured voltage threshold and, thus, can be used to adjust the trigger thresholds of the first discriminator.
The dark count scans are expected to show plateaus indicating an increasing number of SPADs breaking down.
In Fig. \ref{fig:fig011}, these plateaus are visible in all curves acquired for each channel at the three benchmark settings.
The acquired curves for the dark count rates of each channel are compressed for a higher power consumption and stretched out for a lower power consumption.
The dark count rate is slightly increased for $\mathsf{vth\_t1}= 1- 10$ if configuring a higher power consumption (see Fig. \ref{fig:fig011}).
In this configuration, setting  $\mathsf{vth\_t1} > 10$ is sufficient to trigger on a higher number of photo-electrons.
For minimum power consumption, this threshold has to be increased by a factor of 2.5 to reach the same trigger level.

\begin{table*}[b]
	\centering
	\caption{Comparison of power consumption and single-channel performance between different ASIC models.}
	\label{tab:comparison}
	\begin{tabular}{lrrrrrr}
		\hline
		ASIC  & Timestamp Digitization & Charge Measurement & Power Consumption & CRT / ps & Crystal Height  & Ref.\\
		& & &  / mW/channel & &  / mm & \\
		\hline
		FlexToT & external & tot output & 11 & 123 & 5 & \cite{sarasola2017FlexToTvsNINO,comerma2013FlexToT}\\
		HRFlexToT & external & tot comparator & 3.5  & 180 & 20 &\cite{HRFlexTOTManual,gomez2019HRFlexToTNSSMIC}\\
		NINO & external & tot method & 27 & 93 & 5 & \cite{sarasola2017FlexToTvsNINO,anghnolfi2004NINOASIC}\\
		STiC3 & TDC on ASIC & tot method (TDC on ASIC)& 25 & 240 & 15 & \cite{stankova2015STIC3}\\
		PETA4 & TDC on ASIC & qdc method (ADC on ASIC) & $<$ 40 & 460 & 25 & \cite{sacco2013PETA4}\\
		Petiroc & external & tot output & 3.5 (w/o ASIC buffers) & n.a. & n.a. & \cite{fleury2013Petiroc,fleury2017PETIROC2A}\\
		Triroc & TDC on ASIC & qdc method (ADC on ASIC) & 10 & 432.7 & 10 & \cite{ahmad2016Triroc} \\
		TOFPET1 & TDC on ASIC & tot method (TDC on ASIC) & 8 - 11 & 290.7 & 15 & \cite{PETsysTOFPET1DataSheet,niknejad2016TOFPETvalidation}\\
		TOFPET2 & TDC on ASIC & qdc method (ADC on ASIC) & 3.6 - 7.2 (+ 1.2)$^{*}$ & 210 & 5 & \cite{schug2018TOFPET2} \\
		TOFPET2 & TDC on ASIC & qdc method (ADC on ASIC) & 5 - 8 & 202 & 5 & \cite{TOFPET2EKitHardwareGuide,lamprou2018CharacterizationTOFPET}\\
		\hline
		\multicolumn{7}{l}{\scriptsize $^{*}$Values were determined experimentally.}
	\end{tabular}
\end{table*}

\section{Discussion}

The measured power consumption of \SIrange{3.6}{7.2}{\mWperchannel} only includes the power consumption due to the ASIC operation.
An estimate of the power consumption due to the ASIC-FPGA communication can be computed taking an input current of \SI{30}{\milli\ampere} as a reference \cite{PETsysTOFPET2DataSheet}.
Considering a supply voltage of \SI{2.5}{\volt} and 64 ASIC channels, a power consumption of approx. \SI{1.2}{\mWperchannel} has to be added to the measured benchmarks.
The obtained values are in good agreement with the power consumption of \SIrange{5}{8}{\mWperchannel} reported by PETsys Electronics S.A. \cite{TOFPET2EKitHardwareGuide}.\\
Regarding the analytical model, which was implemented to compute the power consumption due to ASIC operation, the determined linearity of 1.034 $\pm$ 0.018 and a negligible y-axis intercept in the order of \SI{0.1}{\mWperchannel} confirm that this model can be used to compute the power consumption prior to experiments and thus to select adequate settings.
The model has only been verified for the present setup and should be tested on different benchtop setups.
The parameter $P_0$ so far can only be determined experimentally.
This parameter correctly accounts for the count rate dependency of the power consumption and thus, also would incorporate effects of changing the discriminator thresholds or applied overvoltage, and employing different SiPM types or scintillator topologies.
It was shown that these changes do not significantly affect the stability of power consumption for a given SiPM configuration in multi-channel experiments.
In addition, the introduced model assumes equal behavior of all parts of the ASIC circuit at each point in the three-dimensional parameter space of ($\mathsf{fe\_ib1}$, $\mathsf{fe\_ib2}$,  $\mathsf{disc\_sf\_bias}$).
Experiments probing the behavior of $\mathsf{fe\_ib1}$ for $\mathsf{fe\_ib2}$ or $\mathsf{disc\_sf\_bias}$ other than zero, as well as related experiments for the other two parameters, should be considered.
However, since the minimum power consumption and various other tuples (see Fig. \ref{fig:fig06}) are correctly described by the model, we do not expect changes to the behavior.\\
Compared to other ASICs the TOFPET2 ASIC features a similar or even lower power consumption.
For other models with similar architecture, higher values are often reported, e.g., \SI{10}{\mWperchannel} for the Triroc ASIC, \SI{25}{\mWperchannel} for the STIC3 ASIC, and less than \SI{40}{\mWperchannel} for the PETA4 ASIC \cite{ahmad2016Triroc,stankova2015STIC3,sacco2013PETA4}.
The prior version of the TOFPET2 ASIC also featured a slightly higher power consumption (\SIrange{8}{11}{\mWperchannel}) \cite{PETsysTOFPET1DataSheet}.
A lower power consumption of \SI{3.5}{\mWperchannel} reported for the Petiroc ASIC does not include the power consumption of the ASIC buffers \cite{fleury2013Petiroc}.
The new version of the FlexToT ASIC, the HRFlexToT ASIC, also comes along with a low power consumption of \SI{3.5}{\mWperchannel} \cite{gomez2019HRFlexToT}.
The Petiroc, the NINO, the FlexToT and the HRFlexToT ASIC do not employ TDCs inside the ASIC circuit, which results in the reported very low power consumptions.
Table \ref{tab:comparison} provides an overview over the given values and the single-channel performance reported along with these.
Due to varying scintillator heights, it is not possible to confirm a clear trend showing a better performance for a higher power consumption.
Comparative studies between different ASIC models, e.g., the FlexToT ASIC (\SI{11}{\mWperchannel}) and the NINO ASIC (\SI{27}{\mWperchannel}) \cite{sarasola2017FlexToTvsNINO,} show that apart from a comparison of the power consumption, multiple parameters, such as linearity of the energy measurement, timing performance as well as the ease of system integration, have to be considered to choose the digitization circuit for a specific application.\\
The pre-amplifier is a low-impedance current conveyor \cite{francesco2016TOFPET2}.
In current-sensing, a low input stage impedance resulting in fast pulses is beneficial for the timing resolution \cite{bugalho2017ExperimentalResultsTOFPET2}.
In contrast a higher input stage impedance results in the slower and higher rise of the acquired pulses.
This behavior is reflected by the performance scans conducted for different configurations of the power consumption.
The systematic change in energy resolution visible in performance experiments depending on the power consumption configuration is suspected to be due to the changed pulse shape for a different input stage impedance.
The acquired dynamic range of the energy spectra does not call for any gain adjustments.
As the filtered count rate does not change between different settings, the observed shift of the 511-keV peak in the energy value spectra does not lead to a loss of events due to the applied energy filter.
The observed CRT deterioration can be explained referring to the increased input impedance when configuring the ASIC to have a lower power consumption.
Due to a slower rise of the registered SiPM signals, the ASIC triggers on these signals at a later point in time, resulting in worse CRTs.
In addition, a shift of the SiPM operation point due to changing the ASIC configuration is visible in the acquired CRT curves, which do not reach a clear minimum for maximum power consumption (see Fig. \ref{fig:pow_consump_vs_performance_CRT}).
Low-power ASICs found in literature also show a worse performance than other ASIC models with a higher power consumption \cite{zhu2016ASIC}.
However, since the TOFPET2 trigger circuit is capable of rejecting noise events, the TOFPET2 ASIC still delivers CRTs down to \SI{300}{\pico\second} at \SI{3.6}{\mWperchannel} power consumption, which is only a \SI{20}{\pico\second} performance loss compared to the default setting.\\
The dark count scans imply that the acquired dark count rate and, accordingly, the photo-electron trigger level is influenced by the configuration of the power consumption, if the trigger threshold is kept at a constant level.
This matches the observation of a higher satellite peak fraction at lower power consumption.
The level to trigger between the first and second or the second and third photo-electron is shifted to higher thresholds.
One has to keep in mind that the acquired dark count rate is influenced by the intrinsic radioactivity of the scintillator coupled to the SiPM array.
Considering an LYSO activity of \SI{500}{\becquerel\per\cm\cubed} \cite{thiel2008LYSOradioactivity} and assuming that each decay is acquired as valid event, the LYSO contribution to the acquired dark count rate would add up to \SI{0.006}{\kilocountspersecond\per\mm\squared}.
This contribution should only be visible as a constant offset in the acquired dark count rate.
Hence it does not contribute to the observed shift of the photo-electron trigger levels.
The effect of shifted trigger levels is probably based on the entire signal processing chain and so far cannot be attributed to one of the parameters changed.
The shift is stronger visible between the default and minimum setting, where multiple parameters were changed.
It can be assumed that the input stage impedance modifies the voltage pulse height and hence, the photo-electron trigger level, since only the parameter $\mathsf{fe\_ib1}$, i.e., the input stage impedance $\mathsf{R_{IN}}$, was changed between the maximum and default configuration (default $\mathsf{R_{IN}} =$ \SI{27}{\ohm}, maximum $\mathsf{R_{IN}} =$ \SI{10}{\ohm}).

\section{Conclusion}

The TOFPET2 ASIC features a power consumption ranging from \SIrange{3.6}{7.2}{\mWperchannel}. 
Including an estimate of the power consumption due to ASIC-FPGA communication, these values increase to \SIrange{4.8}{8.4}{\mWperchannel} and are to our knowledge low compared to the power consumption of other ASIC models.
The reported values are in good agreement with the specifications made by PETsys Electronics S.A..
We present an analytical model allowing the calculation of the power consumption prior to experiments. 
The power consumption is shown to be stable for a range of overvoltages and various count rates ranging from \SI{1}{\kilocountspersecond} to \SI{100000}{\kilocountspersecond}.
Thus, the power consumption remains sufficiently low under various measurement conditions to consider the ASIC for integration in a PET system. \\
As expected, the input stage impedance and discriminator noise have a significant influence on the ASIC performance. 
Depending on the configuration and applied overvoltages, achieved CRTs can be improved by \SIrange{20}{40}{\pico\second}.
For settings apart from the default setting, the energy resolution is deteriorated by up to \SI{0.5}{\percent} (absolute deterioration).
Configuring a lower power consumption results in a shift of the photo-electron trigger levels over the discriminator threshold range.
Therefore, adjustments of the trigger threshold $\mathsf{vth\_t1}$ applied in performance scans are required.
Combining a low power consumption of about \SI{6.4}{\mWperchannel} with approx. \SI{280}{\pico\second} CRT and approx. \SI{10.5}{\percent} energy resolution at default configuration, the TOFPET2 ASIC stands out as a promising candidate for future system developments.

\section{Outlook}

Investigations regarding the adjustments of the trigger threshold $\mathsf{vth\_t1}$ are necessary to deal with the changed photo-electron levels and to provide a fair comparison between the ASIC performance at different power consumption settings.
In addition, a method to separate the loads on the 2.5-V-line needs to be developed to measure the power consumption due to the ASIC-FPGA communication and verify the given estimate.
Since the TOFPET2 ASIC shows not only promising performance, but also low power consumption, it will be further favored for building an MR-compatible TOF-PET insert.
To evaluate the MR-compatibility of the TOFPET2 ASIC we propose similar test protocols as applied in \cite{wehner2014InvestigationOfMRcompatibility,weissler2015MRtests}.

\section{Acknowledgments}

We thank Ricardo Bugalho and Luis Ferramacho from PETsys Electronics S.A. for kindly answering our many questions.

\ifCLASSOPTIONcaptionsoff
\newpage
\fi
\renewcommand*{\bibfont}{\footnotesize}
\printbibliography[heading=bibnumbered]

\end{document}